\documentclass[preprint]{aastex}

\begin{document}


\shorttitle{SDSS~J1650+4251: A New Gravitational Lens}
\title{SDSS~1650+4251: A New Gravitational Lens\altaffilmark{1}}






\author{N.~D.~Morgan\altaffilmark{2,4},
        J.~A.~Snyder\altaffilmark{2,3},
        L.~H.~Reens\altaffilmark{2,4}
}



\altaffiltext{1}{Based on observations obtained with the WIYN telescope at the
Kitt Peak National Observatory, National Optical Astronomy Observatories, which 
is operated by the Association of Universities for Research in Astronomy, Inc. 
(AURA) under cooperative agreement with the National Science Foundation.
The WIYN Observatory is a joint facility of the University of Wisconsin-Madison, 
Indiana University, Yale University, and the National Optical Astronomy Observatories.}

\altaffiltext{2}{Center for Astronomy \& Astrophysics, Yale University, 
New Haven CT 06520; nicholas.morgan@yale.edu, lreens@optonline.net}

\altaffiltext{3}{Also Department of Physics and Department of Astronomy, Yale University, New Haven CT 06520;
jeffrey.snyder@yale.edu}

\altaffiltext{4}{Visiting Astronomer, Kitt Peak National Observatory.}


\begin{abstract}

We report that the Sloan Digital Sky Survey quasar SDSS~J165043.44+425149.3
is gravitationally lensed into two images, based on observations obtained with the 
WIYN 3.5~m telescope at the Kitt Peak National Observatory.  
The lensed quasar, at a redshift of $z=1.54$, appears as two images separated by 
1\farcs2 with $B$-band magnitudes of 17.8 and 20.0.  The lensing galaxy is clearly 
detected in $I$-band images obtained in 0\farcs3 seeing, after point spread 
function subtraction of the two quasar images.  A strong metal-line absorption 
system is also identified in the unresolved SDSS spectrum
of the double quasar, suggesting a plausible lens redshift of $z=0.58$.  The $UBRI$ flux 
ratios of the pair vary significantly from 8.5:1 in the blue to 5.4:1 in the red, a 
difference of 0.5 magnitudes, and may indicate the presence of microlensing in one or 
both quasar components.  The predicted differential time delay between quasar images is 
on the order of one month, assuming the intervening absorption system is due to the
lensing galaxy.

\end{abstract}




\section{Introduction}

Gravitationally lensed quasars are useful for a wide range of applications,
such as constraining the cosmological parameters $H_{\rm{o}}$ (Koopmans et al. 2000; 
Kochanek et al. 2002) and $\Lambda$ (Falco et al. 1998; Chae et al. 2002), mapping 
the structure of luminous and dark matter in intermediate redshift galaxies 
(Koopmans \& Treu 2003), and studying the evolution of the lens galaxies themselves 
(Kochanek et al. 2000).  Owing to a number of large-scale surveys for lensed
quasars at both optical and radio wavelengths, there are now well over 70 confirmed systems.
In this paper, we report that the quasar SDSS~J1650+4251 is gravitationally 
lensed into two images, based on observations obtained with the WIYN 3.5~m telescope.
We present the lens discovery data and subsequent detection of the lensing galaxy in \S2, 
assess the SDSS spectrum of the system in \S 3, and discuss the plausible redshift and
potential model of the lensing galaxy in \S4.  Our findings and possible directions for
future work are summarized in \S5.

\section{Observations and Analysis}

To the best of our knowledge,\footnote{Queries to the NRAO VLA Sky Survey radio catalog, 
the ROSAT X-ray database, and the NASA/IPAC Extragalactic Database returned no results within 2\arcmin\ 
of the quasar's coordinates.} SDSS~J1650+4251 (16$^{\scriptsize{\mbox{h}}}$ 50$^{\scriptsize\mbox{m}}$ 43$\fs$5, 
$+42\arcdeg$ 51$\arcmin$ 45$\farcs$0; J2000.0) was first identified as a $z=1.54$ redshift quasar, 
with a blue magnitude of $g = 17.6$, from the first data release of the Sloan Digital Sky Survey 
(SDSS DR1; {\tt http://www.sdss.org/dr1}).  During the nights of 24 and 25 April 2003, approximately 200 DR1 
quasars were reimaged under excellent seeing conditions with the WIYN telescope at 
the Kitt Peak National Observatory, with the goal of identifying multiple image morphology 
indicative of gravitational lensing.  Our snapshots consisted of 30 second $R$-band 
exposures using the Mini-Mosaic direct imager, which has a field of view of 
9\farcm6 square and a gain and readnoise of 1.5 e$^{-}$ ADU$^{-1}$ and 5.5 e$^{-}$,
respectively.  We operated in 2$\times$2 binning mode to help minimize 
the readout time, which provided a scale of 0\farcs282 pixel$^{-1}$.  

For each target, we assigned an {\it a priori} lensing probability based on the
quasar's redshift and $g$-band magnitude.  These crude probabilities, which are computed using an optical depth calculation and a magnification bias
factor (e.g., Kochanek 1996), were used to select targets for reimaging during the observing run.
The {\it a priori} lensing probability for SDSS~J1650+4251 was 0.7\%, placing it in the top
2\% of the $\sim$8000 DR1 quasars accessible during the WIYN run.  The discovery snapshot 
of the target immediately revealed it to be double with an image separation of 
1\farcs2 and an $R$-band flux ratio of $\sim5:1$.  Binned followup 
images were obtained in Harris $UBRI$ filters, as well as several unbinned images in $I$-band alone,
and revealed a similar separation and flux ratio in the other bandpasses.  The seeing during the 
discovery and followup images
was exceptional, ranging from 0\farcs6 FWHM in $U$ to 0\farcs3 FWHM in $I$.  A summary
of the WIYN data for SDSS~J1650+4251 is provided in Table~1.  In addition, $UBRI$ images of the Landolt (1992)
standard field PG~1633 were also obtained immediately following the quasar observations.
A 2\farcm5 square subraster of the target and surrounding field is shown in Figure~1.

The images were bias-subtracted and flat-field corrected using twilight sky flats taken 
during the run.  To model the light distribution of the double image, we have simultaneously fit 
two empirical point spread functions (PSFs) to components A (brighter) and B (fainter), for each 
image listed in Table~1, using a Powell minimization routine (Press et al. 1992).  For the 
$BRI$-band images, star \#2 in Figure~1 provided the PSF.  This star dropped 
significantly in brightness toward the blue, so star \#1 served as the $U$-band PSF.  After subtracting the
best-fit models, no significant structure was detected in the residual image for the $UBR$ data.  
The average flux ratio for the two components was 5.4:1 in $R$, and increased toward bluer wavelengths to 
7.3:1 in $B$ and 8.5:1 in $U$.  Relative astrometry and photometry for the two-component models 
are reported in Table~2.  Magnitudes in Table~2 have been calibrated using zeropoints 
(but no color terms) obtained from the PG~1633 standards.

The lack of significant residuals in the $UBR$ data indicates that the system is well-modeled by
two point sources at these wavelengths.  However, significant structure was present at 
the position of the double image after subtracting the two-component model from the unbinned $I$-band data, as seen in panel 
(b) of Figure~2.  The peak of the residual flux is $\sim$15$\sigma$ (where $\sigma$ is the rms noise per pixel), and is peaked roughly 
collinear and between the brighter two components.  This position is what would be expected for a 
simple singular isothermal sphere (SIS) lens model of the system, assuming the two brighter and point-like 
images were lensed components of a single background quasar and the faint residual flux was from a 
foreground lensing galaxy.  

To model the suspected galaxy flux (hereafter, component G), a third component was added to the $I$-band 
models and the simultaneous fits were repeated.  The galaxy was modeled using a circularly-symmetric ``pseudogaussian'' 
light profile (Schechter \& Moore 1993), which falls off slower than a true gaussian at moderate distances 
(several pixels) from the profile core.  The relative fluxes and positions of the three components were 
free to vary.  Panel (c) of Figure~2 shows the stacked residuals from the three-component $I$-band fits, 
and is free from significant structure above 5$\sigma$ at the position of component G.  There are still
significant residuals ($\pm8\sigma$) at the cores of component A and B, but this is most likely an
artifact of fractional pixel shifts of the undersampled PSF.  Panel (d) shows the same fit as in panel (c), 
but with the galaxy left unsubtracted.   Results for the three-component fit are also reported in Table~2.

The $I$-band data place the galaxy closer to component B ($\theta_{BG}$=0\farcs36) than to A 
($\theta_{AG}$=0\farcs87), and offset by 0\farcs15 (1 pixel) to the south-west with respect to 
the quasar AB line.  The average $I$-band flux ratio, 5.75:1, is slightly larger than that found at $R$.  
To estimate the galaxy magnitude in $I$, we have performed aperture photometry on
the unsubtracted galaxy flux for the three-component model, that is, on panel (d) of Figure~3.  
An aperture radius of 0\farcs71 (5 pixels) was used, and yields a calibrated magnitude of $I=20.5$.  
Since no aperture correction was applied, this is likely a lower limit to the true galaxy magnitude.

\section{Spectral Observations}

Unresolved spectra of SDSS~J1650+4251 was obtained with the 2.5~m telescope at 
Apache Point, New Mexico, on 19 June 2001 as part of the standard SDSS spectroscopic followup
of color-selected quasar candidates (Richards et al. 2002).  The spectrum has a wavelength
coverage of approximately 3800 \AA\ to 9200 \AA, a dispersion that ranged from 1 to 2 \AA\ 
pixel$^{-1}$ from blue to red, and a total on-source exposure time of 2700 seconds.  A 
5 \AA\ binned sample of the Sloan spectrum is reproduced in the top panel of Figure~3.

The 50\% seeing disc was $\sim2\arcsec$ during the exposure, ensuring that light from both 
components were well mixed.  The unresolved spectrum is consistent with 
at least one object being a $z=1.541$ quasar, based on gaussian fits to the CIII] and MgII broad 
emission lines.  If the remaining component were a foreground star, then one might expect 
zero-redshift stellar features in the unresolved spectrum.  The expected locations 
of several stellar features are marked by the dashed lines in Figure~3, but none
are readily identified with features in the spectrum.  (The absorption feature at the expected
location of the Na doublet is actually a singlet).  The lack of significant stellar 
absorption lines, together with the absence of other quasar emission lines at a redshift different
than that quoted above, argues for two similarly redshifted quasars.  

There is also evidence for two $z>0$ absorption systems in the spectrum, based on four 
doublet features at approximate observed wavelengths of $\lambda\lambda$3945, 4090, 4415, and 7125 
\AA.  The bluest and reddest absorption doublets lie on top of the CIV and MgII quasar emission 
lines, and are consistent with CIV $\lambda\lambda$1548,1550 (bottom left panel of Figure~2) 
and MgII $\lambda\lambda$2796,2803 at redshifts of $z=1.545$ and $z=1.546$, respectively.  
These features are clearly associated with the quasar.  The remaining two doublets are consistent 
with an intervening absorption system at $z=0.577$ arising from FeII $\lambda\lambda$2587,2600 and 
MgII $\lambda\lambda$ 2796,2803 (bottom right of Figure~2).  Assuming this absorber redshift, a 
weak MgI $\lambda$ 2853 absorption singlet was also identified at the same redshift.  

\section{Discussion}

\subsection{Plausible Redshift of the Lensing Galaxy}

Although the lensing galaxy currently lacks a spectroscopic redshift, a plausible
redshift of $z=0.577$ is suggested by the intervening MgII and FeII absorption 
system present in the quasar spectrum.  One can statistically estimate the redshift
of the lensing galaxy following the approach of Kochanek (1992).  Using the image 
separation of $1\farcs2$ and quasar redshift of $z=1.541$, we find a median lens
redshift of $z=0.59$ with a 1$\sigma$ interval of $0.36<z<0.82$ (assuming
matter and energy densities of 0.3 and 0.7 of the closure density, respectively).  

An additional estimate of the galaxy redshift can be obtained from the observed
$I$-band magnitude.  For a given redshift and assuming an isothermal sphere model for
the lensing potential, the galaxy's velocity dispersion $\sigma$ is related
to the observed image separation $\Delta \theta$ by
\begin{equation}
\Delta \theta = \frac{D_{ls}}{D_{os}}\frac{8\pi\sigma^2}{c^2}
\end{equation}
(Narayan \& Bartelmann 1999), where $D_{ls}$ and $D_{os}$ are angular diameter
distances from the lens to the source, and from the observer to the source,
respectively.  The galaxy's $B$-band luminosity can be estimated from its
velocity dispersion using the Faber-Jackson relationship, 
$L/L\star = (\sigma/\sigma\star)^{\gamma}$, where $L\star$ corresponds to a $B$-band magnitude of $M_\star = -19.7 + 5 \log h$.  
and we adopt $\sigma_\star = 225$ km s$^{-1}$ and $\gamma = 4.0$ appropriate for 
elliptical galaxies.  Finally, the galaxy's apparent magnitude $m$ is given by
\begin{equation}
m_{\rm AB}(\lambda_{\rm obs}) - M_{\rm AB}(\lambda_{\rm rest}) = 5 \log \left( \frac{D_{ol}}{10 {\rm pc}} \right) + 7.5 \log (1+z),
\end{equation}
where $D_{ol}$ is the angular diameter distance from the observer to the lens. For 
calculating $M_{AB}(\lambda_{\rm rest})$, a spectral energy 
distribution (SED) for an early-type galaxy was obtained from 
S.~Lilly~(1997, priv.~comm.), which consisted of interpolated and 
extrapolated values of the SEDs presented by Coleman,~Wu, \&~Weedman~(1980). 
The SED is then normalized to the Faber-Jackson luminosity at 
4400(1~+~$z$) \AA, which yields the predicted AB magnitudes.  To transform back to
standard magnitudes, we use $I$~=~$I_{AB}-0.456$ (Fukugita et al. 1995).

Figure~4 shows the predicted galaxy $I$-band magnitude as a function of redshift (curved solid line), along
with the 1$\sigma$ error in the predicted magnitude from the observed dispersion in the
Faber-Jackson relationship (curved dashed lines; Dressler et al. 1987).  The galaxy appears systematically fainter
as it is placed at higher redshifts, up to $z\sim1.2$, since the velocity dispersion 
required to reproduce the observed image separation becomes systematically smaller.  Beyond $z\sim1.2$,
the required velocity dispersion (and therefore apparent magnitude) increases rapidly.
At the observed galaxy magnitude of $I=20.5$ (horizontal line), the formally best galaxy 
redshift is at either $z=0.49$ or $z=1.39$.  The statistical constraint on the lensing
galaxy redshift (shaded region) supports a redshift under unity, as does the observed metal-line
absorption system (vertical tick).  Taken together, the data favors a lensing galaxy redshift in the range of $0.4<z<0.8$, 
consistent with the absorption system at $z=0.577$.

\subsection{Lens Models}

We have fit a lens model to the system, which is straightforward to
model given the simple image geometry.  The misalignment between components A, B, and G
implies that some asymmetry must be present in the lensing potential.  Our 
basic model is an SIS potential for the lensing galaxy, plus an external shear to 
break the circular degeneracy.  It has a two-dimensional effective lensing 
potential $\phi$ given by 
\begin{equation}
\phi = br - \frac{\gamma}{2} r^2 \cos 2(\theta-\theta_{\gamma}).
\end{equation}
The major axis of the isopotential lines point along $\theta_{\gamma}$ (measured E of N), 
which is also the direction (modulo 180$^{\rm o}$) of the perturbing mass
responsible for the shear.  Note that the sign convention for the shear term
is opposite to that adopted by Wisotzki et al. (2002), whose SIS+shear model 
predicts a perturbing mass at $\pm$90$^{\rm o}$ from $\theta_{\gamma}$.  
In the limit of vanishing shear ($\gamma \rightarrow 0)$, the strength of the 
potential, $b$, corresponds to the Einstein radius of the system, which 
is also half of the observed image separation.  
There are five model parameters (including two for the unknown 
quasar source position), and five constraints (the relative positions of 
components A and B with respect to the lensing galaxy and the A/B flux ratio).  
Thus there are no degrees of freedom and our best fit is characterized by $\chi^2=0$.

Parameter values were obtained by minimizing residuals in the source plane.  
For a given parameter set, each quasar image is traced back from the image 
plane to the source plane using the lens equation, and the optimal model 
parameters are then found by minimizing residuals between the model source 
position and the backward projection of the image positions.

Using the $I$-band positions and flux ratio as constraints, our best-fit model parameters are
$b=0\farcs665$, $\gamma=0.152$, and $\theta_{\gamma}=136.8\degr$, and a lensed source
position of ($\Delta \alpha, \Delta \delta$) = (-0\farcs136, 0\farcs218) with respect to the lensing galaxy.  The model predicts 
magnification factors of 4.94 and -0.86 for components A and B, where the negative sign denotes 
a parity flip, and a differential time delay between quasar images of 33.8 days 
for a Hubble constant of $H_{\rm{o}} = 75$ km s$^{-1}$ and assumed galaxy redshift of $z=0.577$.  
At this redshift, the corresponding velocity dispersion 
for the lensing galaxy would be 185 km s$^{-1}$ for an elliptical galaxy.  The Faber-Jackson 
relationship predicts a $B$-band galaxy luminosity of roughly 0.5 L$^{\star}$, which is not 
unreasonable.

The shear strength is fairly large for a double system.  If the source of 
shear was from a neighboring galaxy with the same velocity dispersion and 
redshift as the main lensing galaxy, then it would be located 2\farcs2 away.  
Indeed, there is a possible companion galaxy 3\farcs0 south from the 
lensing galaxy (object G2 in Figure~2), however the position angle of the nearby galaxy
(171.5\degr\ E of N) differs by 35\degr\ degrees from the best-fit 
shear direction.  While this may indicate that the companion is actually a 
chance projection, it is hard to dismiss the notion that the 
nearby galaxy and large shear are associated.

In other lensed quasars with large shears, a nearby group or cluster 
of galaxies at moderate distances (10\arcsec\ to 40\arcsec) sometimes acts as 
the source of the shear (Kundic et al. 1997; Schechter et al. 1997; Kneib et al. 2000).  
For SDSS~J1650+4251, there is one faint object 8\arcsec\ from the lensing galaxy at a PA of 110\degr, 
but the light profile is more consistent with a point source than a galaxy.  
Otherwise, there are no obvious galaxy concentrations near the shear position angle.

\section{Conclusions}

We have reported that the $z=1.54$ quasar SDSS~J1650+4251 is gravitationally lensed into a 
1\farcs2 double.  The lensing galaxy is clearly detected in $I$-band after PSF subtraction
of the two quasar images, and is located closer to the fainter quasar image, as expected for
a simple SIS model of the lensing potential.  A strong metal-line absorption system is detected 
in the unresolved spectrum of the quasar pair, suggesting a plausible lens redshift of $z=0.577$.
Taking this as the lensing galaxy redshift, an SIS plus external shear model predicts a differential 
time delay of 33.8 days ($H_{\rm{o}} = 75$ km s$^{-1}$).  Resolved spectra of the lensing galaxy 
will be needed to obtain a confident galaxy redshift and time delay prediction.

The flux ratios for the system show a significant change (0.5 magnitudes) with wavelength, ranging 
from 5.4:1 in $R$ to 7.3:1 in $B$ to 8.5:1 in $U$.  Since gravitational lensing is achromatic, 
one would ideally expect the flux ratio to be constant, but in practice the individual magnifications
are affected by extinction and microlensing from dust and stars in the lensing galaxy.  Since
the flux ratio for SDSS~J1650+4251 increases towards the blue, this does indeed suggest that perhaps one
of these scenarios is correct.  For at least one doubly lensed quasar, SBS~0909+532 (Kochanek et al. 1997),
the flux ratio is known to vary by up to a full magnitude across optical wavelengths, so the 
magnitude of variation observed for SDSS~J1650+4251 is not unprecedented.  Optical monitoring of the 
double quasar, particularly at blue wavelengths, will help determine if microlensing is indeed
significant and if the lensed quasar is sufficiently variable for time delay studies.

\acknowledgments

The authors would like to extend their appreciation to George Will (Kitt Peak night assistant) and 
Charles Corson (WIYN site engineer) for their help with the WIYN snapshot survey.  We would also 
like to thank Charles Bailyn for comments on the manuscript.

\clearpage



\figcaption[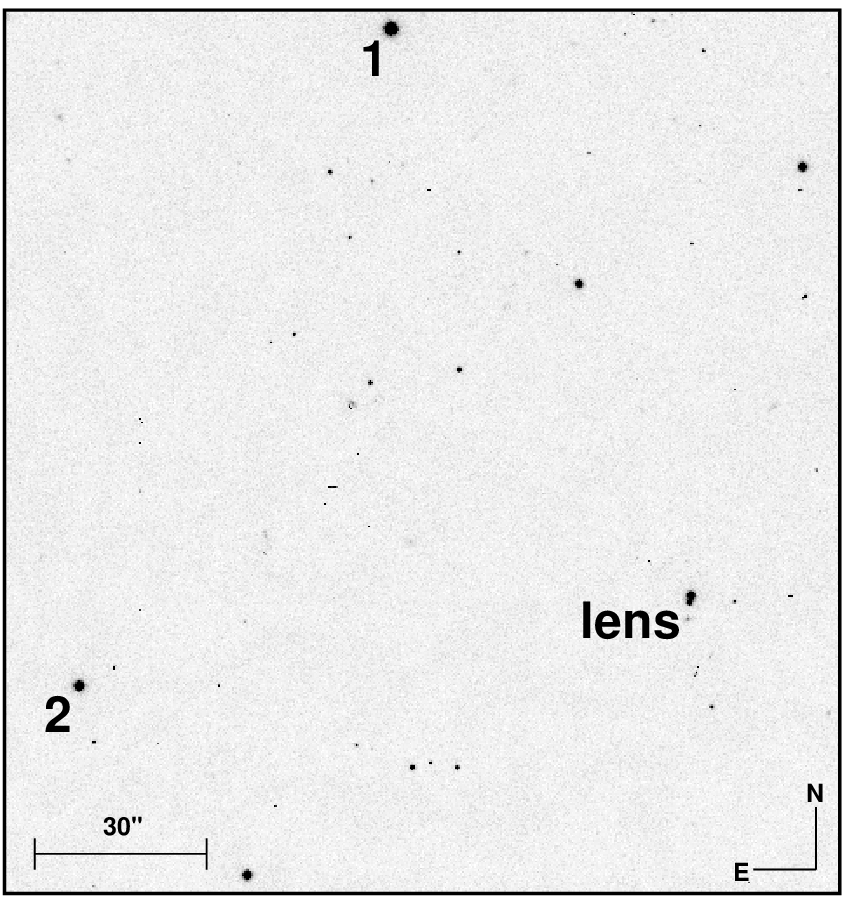] {2\farcm5 square finding chart for SDSS~J1650+4251 and
surroundings.  The lens and two PSF stars are labelled.}

\figcaption[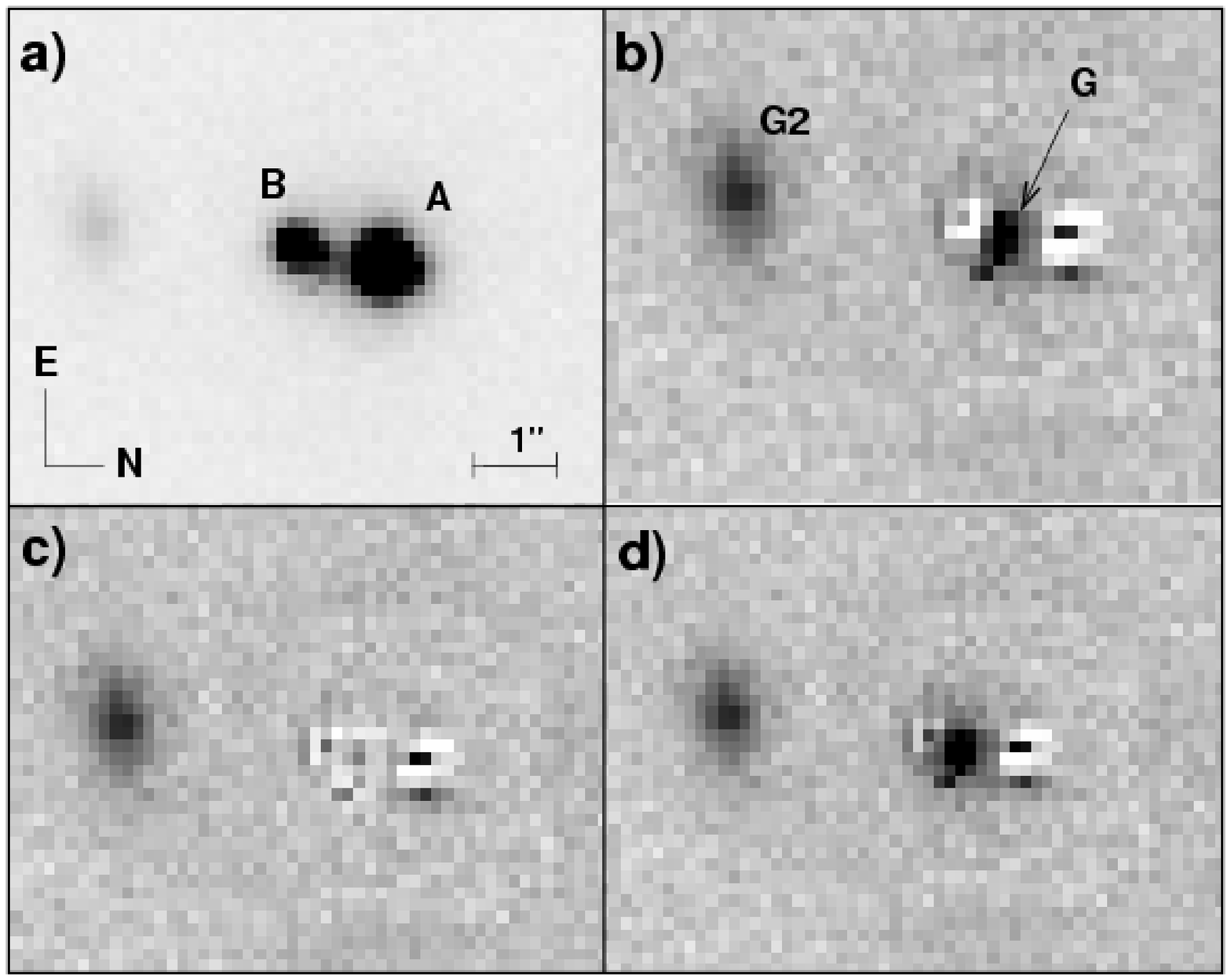] {Subrasters of the SDSS~J1650+4251 taken in 0\farcs3 seeing
with WIYN (scale of 0\farcs141 pixel$^{-1}$).  The orientation and angular size of each panel are 
identical.  {\it Panel (a)}: stacked $I$-band image of the lens.  {\it Panel (b)}: residuals of 
panel (a) after fitting and subtracting two empirical PSFs to components A and B.    
{\it Panel (c)}: residuals of panel (a) after fitting and subtracting two empirical PSFs plus
circularly-symmetric pseudogaussian profile to model the lensing galaxy (component G).  {\it Panel (d)}: 
The same as panel (c), but with the galaxy model unsubtracted.  The contrast level for panels 
(b), (c), and (d) range from -3$\sigma$ to 10$\sigma$, where $\sigma$ is the rms noise 
(Poisson + readout) associated with each pixel.  Note the possible companion galaxy (G2) 3\arcsec\ south 
of the main lensing galaxy.
}

\figcaption[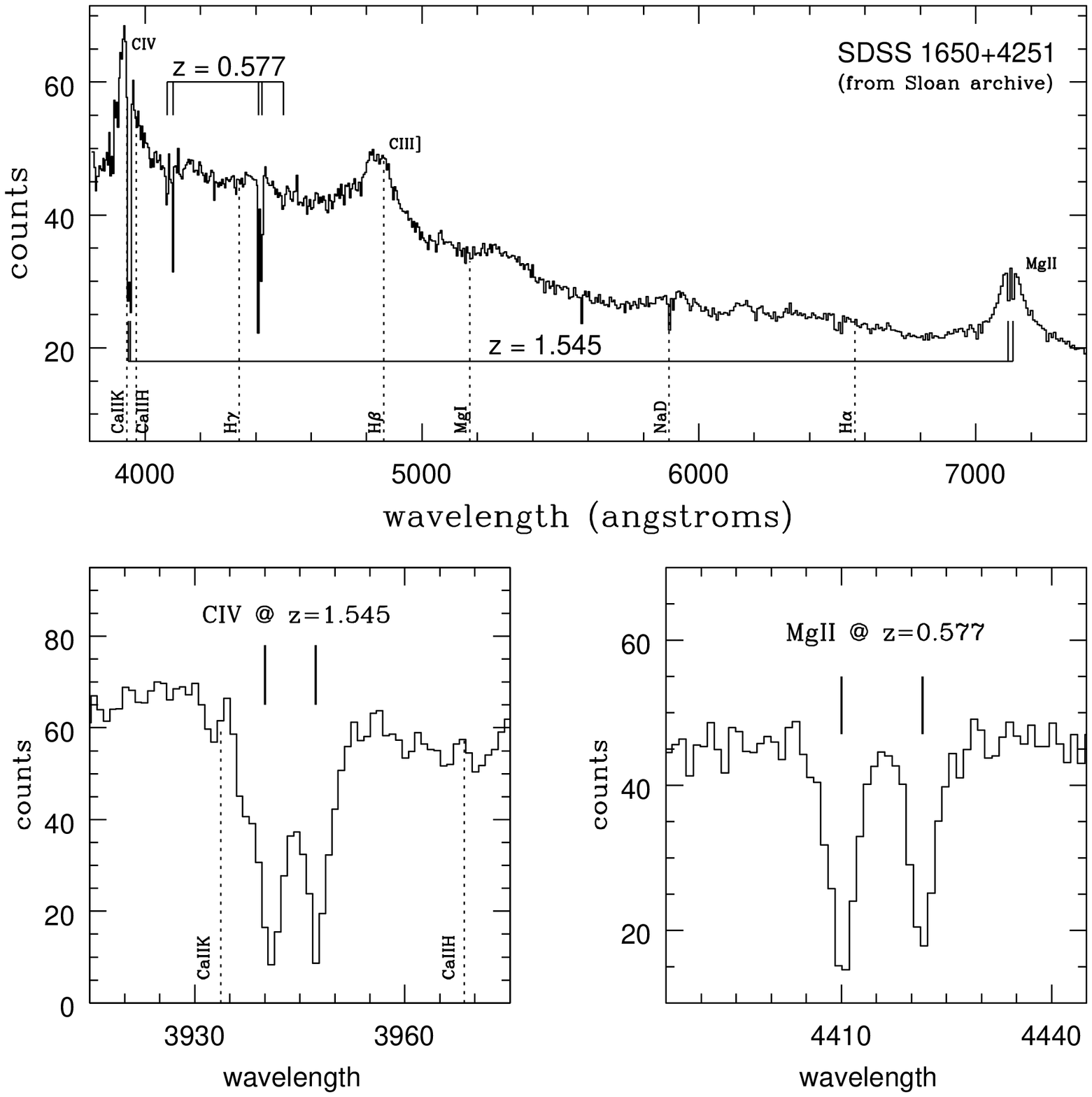] {{\it Top panel}: archive SDSS spectrum of SDSS~J1650+4251.  The
presence of two metal-line absorption systems, one at $z=0.577$ and the other close to the
quasar redshift at $z=1.545$, are indicated by the tick marks.  Dashed lines mark the expected
location of zero-redshift stellar features.  {\it Bottom-left panel}: Closeup of the CIV absorption
doublet at the quasar redshift.  {\it Bottom-right panel}: Closeup of the intervening MgII absorption
doublet at $z=0.577$.
}

\figcaption[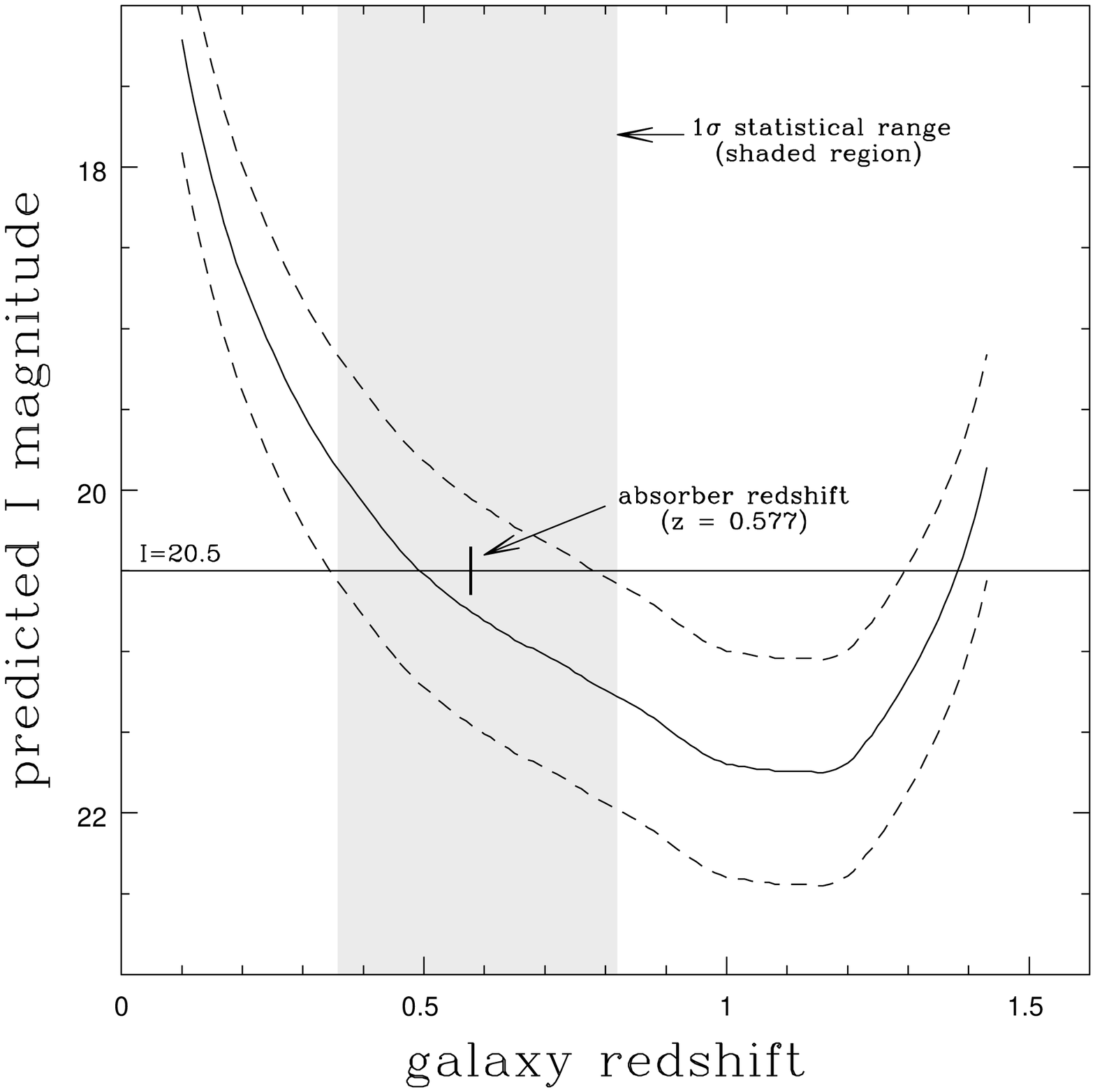] {Predicted $I$-band magnitude of the lensing galaxy as a function
of the lens galaxy redshift (curved solid line), along with the 1$\sigma$ error range (curved 
dashed lines) expected from the scatter in the Faber-Jackson relationship.  The horizontal line
marks the observed $I$-band magnitude of the lensing galaxy ($I$ = 20.5), and the vertical tick marks the 
location of the intervening absorption system ($z=0.577$).  The shaded region ($0.36<z<0.82$) is the $1\sigma$ statistical 
redshift range for the lensing galaxy.
}






\clearpage



 

\makeatletter
\def\jnl@aj{AJ}
\ifx\revtex@jnl\jnl@aj\let\tablebreak=\nl\fi
\makeatother


\begin{deluxetable}{ccccc}
\tablecaption{Log of Observations for SDSS~1650+4251
\label{TABLE1}}
\tablenum{1}
\tablewidth{0pt}
\tablehead{  
\colhead {Frame} &
\colhead {Filter} &
\colhead {Exp. (s)}&
\colhead {FWHM (\arcsec)} &
\colhead {Bin Mode}
}
\startdata
302 & R & 30  & 0\farcs41 & 2 \nl
304 & U & 60  & 0\farcs58 & 2 \nl
305 & U & 90  & 0\farcs60 & 2 \nl
306 & I & 60  & 0\farcs38 & 2 \nl
307 & B & 60  & 0\farcs45 & 2 \nl
308 & I & 60  & 0\farcs32 & 1 \nl
309 & I & 150 & 0\farcs31 & 1 \nl
310 & I & 150 & 0\farcs32 & 1 \nl
311 & I & 150 & 0\farcs33 & 1 \nl
313 & R & 60  & 0\farcs40 & 2 \nl
314 & B & 60  & 0\farcs47 & 2 \nl
\enddata
\tablecomments{Binning modes 1 \& 2 yield a pixel scale of 0\farcs141 pixel$^{-1}$ \& 0\farcs282 pixel$^{-1}$, respectively.}
\end{deluxetable}

\clearpage



 

\makeatletter
\def\jnl@aj{AJ}
\ifx\revtex@jnl\jnl@aj\let\tablebreak=\nl\fi
\makeatother


\begin{deluxetable}{cccccccc}
\tablecaption{Astrometry and Photometry for SDSS~1620+4251
\label{TABLE2}}
\tablenum{2}
\tablewidth{0pt}
\tablehead{  
\multicolumn{4}{c}{ } &
\multicolumn{2}{c}{ B } & 
\multicolumn{2}{c}{ G } \\
\multicolumn{4}{c}{ } &
\multicolumn{2}{c}{ -------------------------------- } & 
\multicolumn{2}{c}{ -------------------------------- } \\
\colhead {Filter} &
\colhead {$m_A$} &
\colhead {$m_B$} &
\colhead {$m_G$} &
\colhead {$\Delta$R.A. (\arcsec)} &
\colhead {$\Delta$Dec. (\arcsec)} &
\colhead {$\Delta$R.A. (\arcsec)} &
\colhead {$\Delta$Dec. (\arcsec)}
}
\startdata
U & 17.10 & 19.42 & --- & 0.212 \phantom{$\pm$ 0.000} & -1.138 \phantom{$\pm$ 0.000} & --- & --- \nl
B & 17.80 & 19.96 & --- & 0.212 \phantom{$\pm$ 0.000} & -1.145 \phantom{$\pm$ 0.000} & --- & --- \nl
R & 17.44 & 19.26 & --- & 0.223 \phantom{$\pm$ 0.000} & -1.147 \phantom{$\pm$ 0.000} & --- & --- \nl
I & 17.15 & 19.05 & $20.5$ & 0.223 $\pm$ 0.002           & -1.163 $\pm$ 0.001           & 0.017 $\pm$ 0.032 & -0.872 $\pm$ 0.026 \nl
\enddata
\tablecomments{Harris magnitudes for components A, B, and G are denoted by 
$m_A$, $m_B$, and $m_G$, respectively.  Relative astrometry for components
B and G are with respect to component A.  Error bars for the $I$-band positions are from the rms dispersion among the four unbinned $I$-band frames.}
\end{deluxetable}

\clearpage


\thispagestyle{empty}
\begin{figure}[h]
\vspace{7.0 truein}
\includegraphics{NickMorgan.fig1.ps}
\end{figure}
\clearpage

\thispagestyle{empty}
\begin{figure}[h]
\vspace{7.0 truein}
\includegraphics{NickMorgan.fig2.ps}
\end{figure}
\clearpage

\thispagestyle{empty}
\begin{figure}[h]
\vspace{7.0 truein}
\includegraphics{NickMorgan.fig3.ps}
\end{figure}
\clearpage

\thispagestyle{empty}
\begin{figure}[h]
\vspace{7.0 truein}
\includegraphics{NickMorgan.fig4.ps}
\end{figure}
\clearpage


\end{document}